\documentclass[prc,twocolumn,
showpacs,amsmath,amssymb,floatfix,nofootinbib]{revtex4-1}
\pdfoutput=1
\usepackage{graphicx,bm,url,braket,xcolor,xspace}
\usepackage{hyperref}
\usepackage[normalem]{ulem}
\newcommand{\fmi}{\, \text{fm}^{-1}}

\newcommand{\bb}{$\beta\beta$\xspace}
\newcommand{\bbz}{$0\nu\beta\beta$\xspace}
\newcommand{\bbt}{$2\nu\beta\beta$\xspace}
\newcommand{\vlk}{$V_{{\rm low}\,k}$\xspace}
\newcommand{\Se}{$^{82}$Se\xspace}
\newcommand{\Ge}{$^{76}$Ge\xspace}

\newcommand{\mev}{\, \text{MeV}}
\newcommand{\Q}{\hat Q}
\newcommand{\X}{\hat X}
\newcommand{\hw}{\hbar\omega}

\begin{document}

\title{Effective double-beta-decay operator for $^{76}$Ge and $^{82}$Se}

\author{Jason D.\ Holt}
\affiliation{Institut f\"ur Kernphysik, Technische Universit\"at Darmstadt, 64289 
Darmstadt, Germany} 
\affiliation{ExtreMe Matter Institute EMMI, GSI Helmholtzzentrum f\"ur 
Schwerionenforschung GmbH, 64291 Darmstadt, Germany}
\affiliation{Department of Physics and Astronomy, University of Tennessee,
Knoxville, TN 37996, USA}
\affiliation{ Physics Division, Oak Ridge National Laboratory, P.O. Box 2008,
Oak Ridge, TN 37831, USA}
\email{jason.holt@physik.tu-darmstadt.de}

\author{Jonathan Engel}
\affiliation{Deptartment of Physics and Astronomy, University of North Carolina,
Chapel Hill, NC, 27516-3255, USA}
\email{engelj@physics.unc.edu}

\begin{abstract}
We use diagrammatic many-body perturbation theory in combination with
low-momentum interactions derived from chiral effective field theory to
construct effective shell-model transition operators for the neutrinoless
double-beta decay of \Ge and \Se.  We include all unfolded diagrams to first-
and second-order in the interaction and all singly folded diagrams that can be
constructed from them.  The resulting effective operator, which accounts for
physics outside the shell-model space, increases the nuclear matrix element by
about 20\% in \Ge and 30\% in \Se.
\end{abstract}
 \date{\today}
\maketitle

\section{\label{s:introduction}Introduction}

The experimental discovery of neutrinoless double-beta (\bbz) decay, a
nuclear-weak process that occurs extremely slowly if at all, would have deep
implications for particle physics.  Since \bbz decay can occur only if the
neutrino is its own antiparticle, an observation would at once establish the
neutrino as a Majorana particle.  Furthermore, from a measured lifetime we
could, in the absence of exotic new physics, determine an average neutrino mass
$m_\nu \equiv \sum_i U_{ei} m_i^2$, where $i$ labels the mass eigenstates, and
$U$ is the neutrino mixing matrix~\cite{avi08}.  This mass cannot be extracted
from a lifetime, however, without first knowing the value of a nuclear matrix
element that also plays a role in the decay. The entanglement of nuclear and
neutrino physics has led to a small but concentrated effort within the nuclear
structure community to calculate the nuclear matrix elements, which are not
themselves observable.  While various theoretical approaches agree to within
factors of two or three, --- a range many structure theorists might find not
unreasonable --- the uncertainty in the effective mass that can be extracted
from an observed lifetime is at least that large as a result.  Since
large-scale experiments will be reporting results in the coming years, we need
to work quickly to improve the accuracy of the matrix-element calculations.

Of the theoretical methods currently employed, the nuclear shell model is the
only approach that offers an exact treatment of many-body correlations, albeit
within a truncated single-particle (valence) space above some assumed inert
core.  Though most of the physics governing double-beta (\bb) decay indeed
resides in this valence space, correlations involving neglected single-particle
orbitals may contribute non-negligibly to both the Hamiltonian and the
\bbz-decay transition operator, each of which is a basic ingredient in any
nuclear matrix-element calculation.  Contributions to the Hamiltonian can be,
and have been, included in the construction of an {\it effective} valence-space
Hamiltonian, $H_{\rm eff}$, through diagrammatic many-body perturbation theory
(MBPT)~\cite{bra67}.  But the analogous contributions to an effective
valence-space \bbz-decay operator, with the exception of a crude
renormalization of $g_A$, have thus far been almost completely ignored.  The
first and only work to apply MBPT to the \bbz-decay operator considered only
diagrams that were first order in the interaction (a $G$-matrix), plus a few
selected higher-order contributions~\cite{eng09}.

In this article we carry out a much more comprehensive computation, providing
the first steps towards a true first-principles calculation of nuclear matrix
elements based on chiral nuclear forces~\cite{chiral}.  We first define and
compute an $\X$-box consisting of all diagrams now to second order in the
interaction and in a much larger Hilbert space than used in Ref.~\cite{eng09}.
We then consider contributions of folded diagrams together with state norms,
which must be explicitly computed for effective transition operators (see
Eq.~\eqref{eq:folding}).  We finally apply the resulting two-body effective
\bbz-decay operator, together with wavefunctions from existing shell-model
calculations, to obtain corrected nuclear matrix elements in the $pf$-shell
\bbz-decay candidates \Ge and \Se . 

Assessing the accuracy of the perturbative expansion is a central challenge for
MBPT.  Although a nonperturbative treatment of core polarization found only
modest changes in $H_{\rm eff}$~\cite{hol05}, the analogous impact on effective
\bb-decay operators is unclear, and ultimately a nonperturbative method that
goes beyond core polarization, allowing controlled approximations to both the
effective Hamiltonian and transition operators, will be preferable.
Coupled-cluster theory~\cite{hagen10,jansen11} and the in-medium similarity
renormalization group~\cite{tsukiyama10,tsukiyama12,hergert13} are promising
nonperturbative methods, but neither has yet been applied to \bb decay.  The
situation may be different in a few years, but at present MBPT is still the
best method to investigate microscopic many-body corrections to the shell-model
\bbz-decay operator.  And even within MBPT, as we have noted, there is
essentially no work, outside of Ref.\ \cite{eng09}, on two-body transition
operators, making the topic almost completely unexplored.

The remainder of this paper is structured as follows: Section \ref{s:methods}
describes the ingredients of our calculation, including definitions of the
matrix elements we compute, the framework for obtaining the nuclear
interactions with which we begin, and the details of our many-body formalism
for calculating effective \bb-decay operators.  Section \ref{s:results}
presents our results for \Ge and \Se, updating the matrix element for \Se first
reported in Ref.~\cite{lin01}.  Finally, Section \ref{s:discussion} discusses
the significance of the results and outlines steps that will improve their
accuracy.

\section{\label{s:methods}Methods}

\subsection{Decay Operator}

In the closure approximation (which is good to at worst 10\% or so
\cite{pan90}), the nuclear matrix element governing \bbz decay can be
represented as the ground-state-to-ground-state matrix element of a two-body
operator.  Neglecting the so-called ``tensor term,'' the effect of which is a
few percent~\cite{kor07a,men08}, the matrix element is given by
\begin{align}
\label{eq:me}
 M_{0\nu} & =  \frac{2R}{\pi g_A^2} \int_0^\infty \!\!\! q \, dq  \\ 
 &\times \bra{f} \sum_{a,b}\frac{j_0(qr_{ab})\left[ h_F(q)+   
 h_{GT}(q) \vec{\sigma}_a \cdot \vec{\sigma}_b \right]}
 {q+\overline{E}-(E_i+E_f)/2} \tau^+_a \tau^+_b \nonumber \ket{i} \,,
\end{align}
where $\ket{i}$ and $\ket{f}$ are the ground states of the initial and final
nuclei, $r_{ab} \equiv |\vec{r}_a-\vec{r}_b|$ is the distance between nucleons
$a$ and $b$, $j_0$ is the usual spherical Bessel function, and the nuclear
radius $R$ is inserted to make the matrix element dimensionless, with a
compensating factor in the phase-space integral that multiplies the matrix
element.  The ``form factors'' $h_F$ and $h_{GT}$ are given by
\begin{align}
\label{eq:form-facs}
h_F(q) &\equiv -g_V^2(q^2) \,,\\
h_{GT}(q)&\equiv g_A^2(q^2) -\frac{g_A(q^2) g_P(q^2) q^2}{3m_p} +
\frac{g_P^2(q^2)q^4}{12m_p^2}  \nonumber \\
&+ \frac{g_M^2(q^2)q^2}{6m_p^2} \,, \nonumber 
\end{align}
where
\begin{align}
\label{eq:dipoles1}
g_V(q^2)&=\frac{1}{\left(1+q^2/(0.85 \textrm{GeV}^2)\right)^2}\,, \\
g_A(q^2)&=\frac{1.27}{\left(1+q^2/(1.09 \textrm{Gev}^2)\right)^2} \,, \nonumber
\end{align}
\begin{align}
\label{eq:dipoles2}
g_P(q^2)&=\frac{2m_p g_A(q^2)}{q^2+m_\pi^2} \,,  \qquad g_M(q^2) = 3.70
g_V(q^2) \,,  \nonumber
\end{align}
and $m_p$ denotes the proton mass and $m_\pi$ the pion mass.

The closure approximation is not good for two-neutrino double-beta (\bbt)
decay, which we briefly discuss later. The matrix element governing that
process contains a complete set of intermediate states, viz.:
\begin{equation}
\label{eq:two-nu}
M_{2\nu}  \approx \sum_n \frac{\bra{f}\sum_a \vec{\sigma}_a \tau^+_a
\ket{n} \bra{n}\sum_b
\vec{\sigma}_b \tau^+_b \ket{i}}{E_n-(M_i+M_f)/2} \,,
\end{equation}
where $n$ denotes states in the intermediate nucleus with energy $E_n$, $M_i$
and $M_f$ are the masses of the initial and final nuclei, and the effects we
neglect (e.g., forbidden currents, the Fermi matrix element, etc.) are small.
We are unable to obtain a complete set of intermediate states, so we can treat
\bbt decay only in the closure approximation, viz.:
\begin{equation}
\label{eq:two-nu-cl}
M_{2\nu}^\text{cl}  = \bra{f}\sum_{ab} \vec{\sigma}_a \cdot \vec{\sigma}_b
\tau^+_a \tau^+_b \ket{i} \,.
\end{equation}
Although the approximation is poor, and we cannot use it to deduce the real
\bbt-decay matrix element, the closure matrix element and the real one change
in a similar way when correlations are added.

\subsection{Nuclear Interactions}

Diagrammatic MBPT was reviewed extensively some years ago \cite{kuo91,hjo95},
but since then, driven by advances in chiral effective field theory
(EFT)~\cite{chiral} and renormalization-group (RG) methods~\cite{vlowk}, it has
seen something of a revival~\cite{cor09}.  Chiral EFT is a systematic expansion
of nuclear interactions and electroweak currents in which three- (3N) and
higher-body forces arise naturally.  Beginning from the chiral two-nucleon (NN)
potential of Ref.~\cite{n3lo}, we construct a low-momentum interaction (\vlk),
with cutoff $\Lambda=2.0\fmi$, via RG evolution~\cite{vlowk,smooth}, explicitly
decoupling high-momentum components from those at the nuclear-structure scale.
In contrast, the $G$-matrix~\cite{hjo95}, often taken as a starting point in
nuclear structure calculations, deals with high-momentum modes by
particle-ladder resummation, and does not adequately decouple low- from
high-momentum degrees of freedom.  As a result, many-body methods based on \vlk
tend to converge better than those using a $G$-matrix~\cite{hagen10}. Recent
work with MBPT based on low-momentum NN+3N interactions has led to the
development of non-empirical valence-space Hamiltonians for proton- and
neutron-rich systems~\cite{ots10,hol13a,hol12,gal12,hol13b}.  While 3N forces
are neglected here, we plan to include them in our future \bbz-decay
nuclear-matrix-element calculations.

The one drawback of using low-momentum interactions in calculations of
effective operators is that high-momentum physics cannot be included
explicitly.  The effects of high-momentum (short-range) correlations on the
\bbz-decay operator are both small and now well understood, however, and we
include them via an effective Jastrow function that has been fit to the results
of Brueckner-theory calculations~\cite{sim09}. 

\subsection{Effective Two-Body Transition Operators}

As we have noted, existing work on MBPT contains little about effective
two-body operators other than the Hamiltonian, where
Refs.~\cite{bra67,kre75,ell75} provide the most comprehensive discussion. No
matter the two-body operator of interest, however, the starting point is always
the construction of projection operators $\hat{P}$ and $\Q$ that divide the
full many-body Hilbert space into a model space, in which subsequent exact
diagonalization is carried out, and everything else.  In our calculations in
nuclei with mass near $A=80$, the model space consists of the $0f_{5/2}$,
$1p_{3/2}$, $1p_{1/2}$, and $0g_{9/2}$ single-particle orbits, for both protons
and neutrons, above a $^{56}$Ni core in a harmonic-oscillator basis of 13 major
shells with $\hbar\omega=10.0\mev$.

\begin{figure}[t]
\centering
\includegraphics[height=7.5cm]{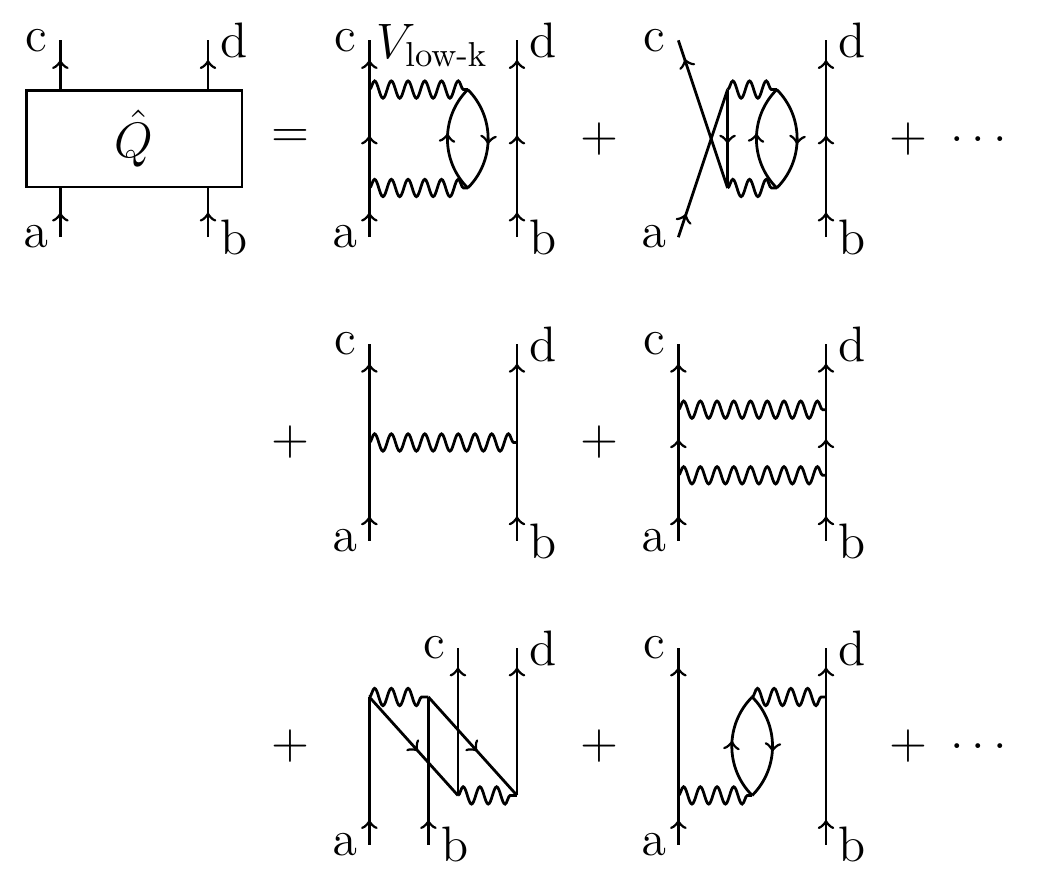}
\caption{\label{fig:qbox} The $\Q$-box to second order in \vlk (ellipses
indicate higher-order terms).  The first line contains one-body contributions
and the others two-body contributions.  Exchange diagrams, though not shown,
are included in our calculations.}
\end{figure}

\begin{figure}[b]
\centering
\includegraphics[height=7.5cm]{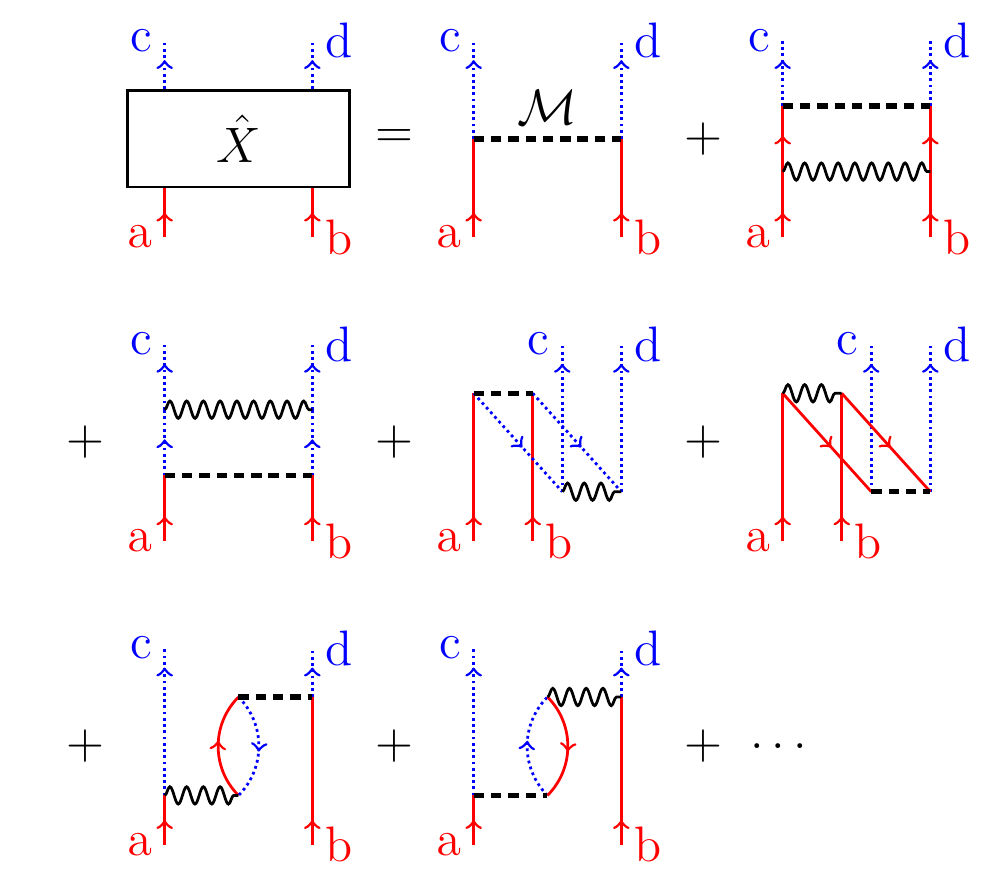}
\caption{\label{fig:xbox} (Color online) The $\X$-box to first order in \vlk.
Solid (red) up- or down-going lines indicate neutrons and dotted (blue) lines
protons.  The wavy horizontal lines, as in Fig.\ \ref{fig:qbox}, represent
\vlk, and the dashed horizontal lines represent the \bbz-decay operator in
Eq.~\eqref{eq:me}.}
\end{figure}

After specifying the model space, one must define a mapping between eigenstates
of the full Hamiltonian and projections of those eigenstates onto the model
space.  In MBPT this is done perturbatively. The result is a set of diagrams
with two incoming legs and two outgoing legs, with each diagram representing a
contribution to the two-body matrix elements of the effective Hamiltonian or
effective (two-body) transition operator. The usual Feynman rules are used to
evaluate the diagrams, but to the set of familiar-looking diagrams one must add
``folded'' diagrams, which eliminate the energy dependence of the effective
operator~\cite{kuo91,hjo95}.  One way to organize the sum of all diagrams is by
grouping all those without folds into a ``$\Q$-box'' (for the Hamiltonian) or
an ``$\X$-box'' (for the transition operator) and then writing the complete
sum, including folded diagrams, in terms of the $\Q$- and $\X$-boxes and their
derivatives with respect to unperturbed energies.  The first few terms in the
$\Q$- and $\X$-boxes appear in Figs.~\ref{fig:qbox} and~\ref{fig:xbox}.

\begin{figure*}[t!]
\centering
\includegraphics[width=.8\textwidth,height=.42\textheight]{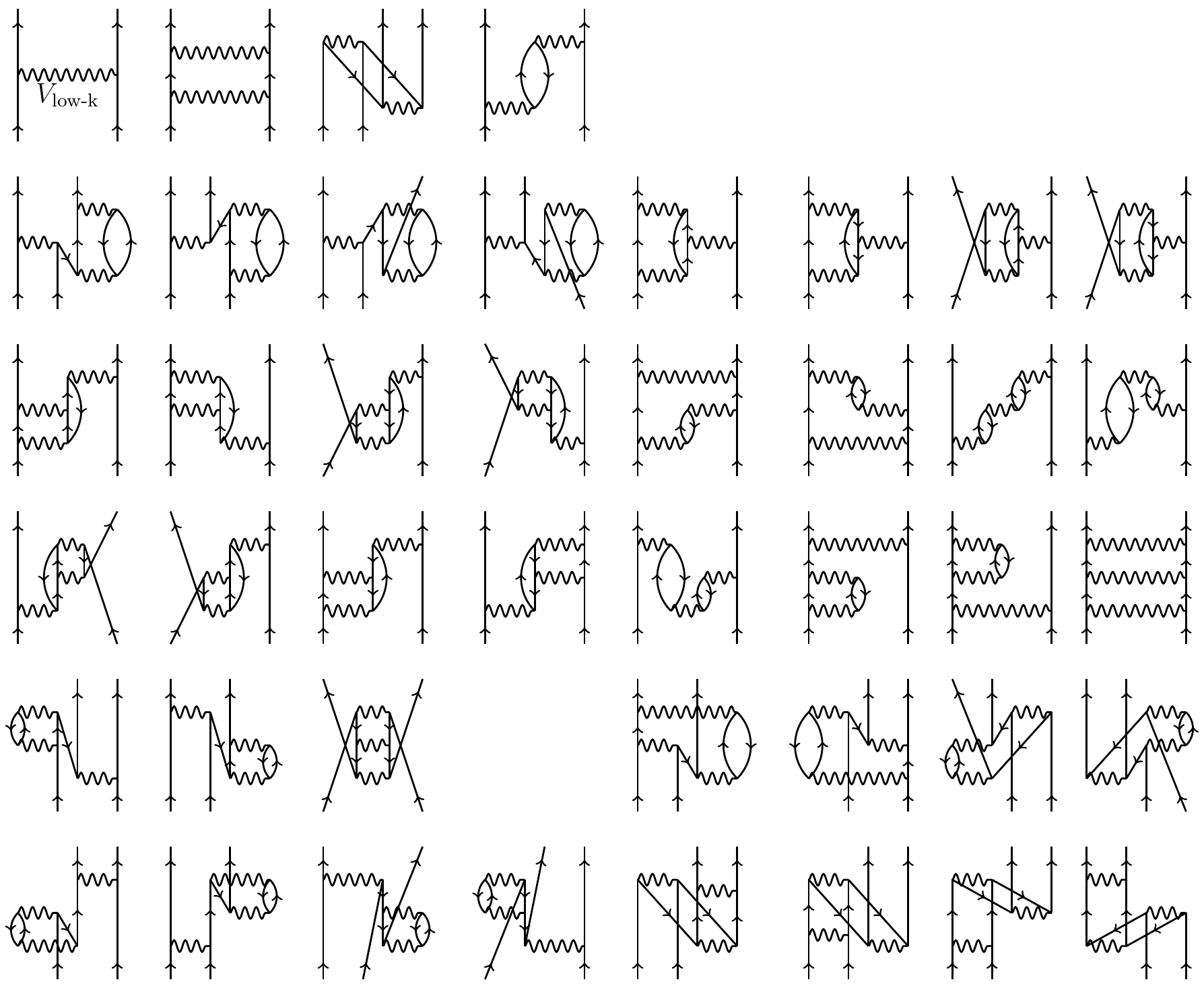}
\caption{\label{fig:third-order} Diagrams in the expansion of the effective
interaction defining the two-body part of the second- and third-order
$\Q$-box.  The wavy lines represent \vlk. We obtain the first- and second-order
$\X$-box --- the set of all unfolded first- and second-order diagrams for the
two-body effective operator (not including norm diagrams) --- by replacing one
interaction in each of these diagrams by a \bb-decay operator (in all possible
ways) and restricting the sums over nucleons in the intermediate states to
either neutrons or protons, as in the first-order $\X$-box diagrams in
Fig.~\ref{fig:xbox}.}
\end{figure*}

Folding is significantly more complicated for a two-body transition operator,
which combines $\X$- and $\Q$-boxes, than for the Hamiltonian, where only
$\Q$-boxes are needed.  Effective model-space operators in the basis of energy
eigenstates are always defined (for a bare operator $\mathcal{M}$) via 
\begin{equation} 
\label{eq:oef}
\frac{\bra{f_\text{eff}} \mathcal{M}_\text{eff}
\ket{i_\text{eff}}}{\braket{f_\text{eff}|f_\text{eff}}^{\frac{1}{2}}
\braket{i_\text{eff}|i_\text{eff}}^{\frac{1}{2}}} =
\bra{f}\mathcal{M}\ket{i} \,,
\end{equation}
where the states that lie in the model space, $\ket{i_\text{eff}} \equiv
\hat{P} \ket{i}$ and $\ket{f_\text{eff}} \equiv \hat{P} \ket{f}$, are not in
general normalized.  If $\mathcal{M}$ is the Hamiltonian, then only diagonal
matrix elements are nonzero, and the denominator is canceled by a similar
factor in the numerator.  For two-body transition operators, that is not the
case, and state norms must be explicitly computed.  Prior authors have
approached the issue of norms in several ways.  References \cite{bra67} and
\cite{ell75}, for instance, choose to expand the denominators and fold them
into the numerators, thus completely eliminating all disconnected diagrams. The
resulting expressions, however, become complicated as the number of folds
increases, and the approach requires the construction of a special basis as an
intermediate step.  For these reasons Ref.~\cite{kre75} advocates keeping the
denominator and numerator separate, at the price of introducing disconnected
diagrams that only cancel when the sum is carried out completely.  Here, though
we evaluate the $\Q$-box to third order and the $\X$-box to second order in the
interaction, we include only one fold in each of the three factors on the left
hand side of Eq.~\eqref{eq:oef}, and so opt to follow Refs.~\cite{bra67,kre75}
in expanding the denominator and folding with the numerator.  The resulting
expression for the matrix elements of an operator $\mathcal{M}_\text{eff}$ is
approximately\footnote{Because off the need for a special basis, this
expression is only strictly correct when the terms in square brackets are
diagonal.  They are close to diagonal in the calculations presented here.}
\begin{eqnarray}
\label{eq:folding}
&&\bra{cd}\mathcal{M}_\text{eff}\ket{ab} = \\
&& \left( \left[
1+\frac{1}{2}\frac{d\hat{Q}(\varepsilon)}{d\varepsilon}+
\frac{1}{2}\frac{d^2\hat{Q}(\varepsilon)}{d^2\varepsilon}\hat{Q}(\varepsilon) +
\frac{3}{8} \left( \frac{d\hat{Q}(\varepsilon)}{d\varepsilon} \right)^2  
 \ldots \right] \right. \nonumber \\ 
 &&\times \left[ \hat{X}(\varepsilon) + \hat{Q}(\varepsilon)
\frac{\partial \hat{X}(\varepsilon_f,\varepsilon)}{\partial \varepsilon_f}
\bigg|_{\varepsilon_f=\varepsilon}\!\!\!\! + \frac{\partial
\hat{X}(\varepsilon,\varepsilon_i)}{\partial \varepsilon_i}
\bigg|_{\varepsilon_i=\varepsilon} \!\!\! \hat{Q}(\varepsilon)  \ldots \right]
\nonumber \\
&& 
\left. \times \left[
1+\frac{1}{2}\frac{d\hat{Q}(\varepsilon)}{d\varepsilon}+
\frac{1}{2}\frac{d^2\hat{Q}(\varepsilon)}{d^2\varepsilon} \hat{Q}(\varepsilon) +
\frac{3}{8} \left( \frac{d\hat{Q}(\varepsilon)}{d\varepsilon} \right)^2  
 \ldots \right] \right)_{cd,ab}
 \nonumber
\end{eqnarray}
where $\varepsilon$ is the unperturbed energy of both the initial and final
states (we take the energies to be the same).  Both $\Q$ and $\X$ are matrices,
with indices corresponding to the possible two-body states in the valence space
(e.g., $a, b$ or $c, d$ in Figs.~\ref{fig:qbox} and \ref{fig:xbox}).  In this
paper we report results of just the terms explicitly given above, which contain
between zero and five folds (there is a fold at every matrix multiplication).
The terms indicated by ellipses are more complicated and presumably less
important; they await future investigation.

\subsection{\label{ss:xbox}Evaluation of $\Q$- and $\X$-Box Diagrams}

\begin{figure}
\centering
\includegraphics[width=.6\columnwidth]{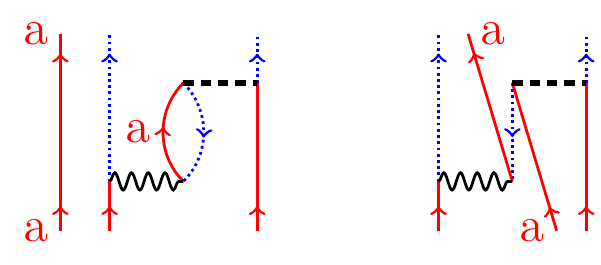}
\caption{\label{fig:cancel} (Color online) A Pauli-forbidden two-body diagram
with a spectator neutron and a three-body diagram, obtained by exchanging two
ingoing neutron lines, that cancels it exactly.}
\end{figure}

We turn now to the $\Q$- and $\X$-boxes themselves, constructed from unfolded
diagrams, that we use in Eq.\ (\ref{eq:folding}).  To construct the $\Q$-box,
we take all unfolded diagrams to third order in \vlk. The diagrams appear in
Appendix A.2 of Ref.~\cite{hjo95}, and the two-body pieces are reproduced in
Fig.~\ref{fig:third-order}.  Our $\X$-box has too many diagrams to display
here, so we characterize the set as follows: we take all two-body $\Q$-box
diagrams in Fig.~\ref{fig:third-order} and replace one interaction line in each
diagram (in all possible ways) by a \bb-decay line.  We then determine whether
each intermediate-state nucleon line should be a proton or neutron. The result
is three times as many $\X$-box diagrams (at second order in \vlk) as $\Q$-box
diagrams in Fig.~\ref{fig:third-order}.

We make one nonstandard choice in evaluating the $\X$-box: we restrict the
particle lines in the intermediate states to be essentially unoccupied.  For
example, in the \bb decay of \Ge, we omit all contributions from intermediate
protons in the $1p_{3/2}$ orbit and neutrons in $1p_{3/2}$, $1p_{1/2}$, or
$0f_{5/2}$ orbits, ad we multiply the contributions of graphs with intermediate
neutrons in the $0g_{9/2}$ orbit by 0.4, its average occupancy.  In the decay
of \Se, we omit the same contributions as in \Ge and multiply the contributions
of graphs with intermediate neutrons in the $0g_{9/2}$ orbit by 0.2 and those
with intermediate protons in the $0f_{5/2}$ orbit by 0.5.  The reason for all
this is that in a nucleus with more than two valence nucleons, the diagram on
the left of Fig.~\ref{fig:cancel} --- a two-body contribution to the \bb-decay
operator with a spectator neutron --- would be canceled by the three-body
diagram on the right if we were to include it. By omitting the two-body
diagrams with intermediate particles in occupied orbits we are effectively
adding particular three-body diagrams (like those on the right of
Fig.~\ref{fig:cancel}) to our calculation.  We are not including all three-body
diagrams, just those that cancel Pauli-forbidden two-body diagrams.

We call this approach nonstandard because it is not usually followed in
derivations of effective interactions.  The reason is that in excluding some
Pauli-forbidden diagrams, one effectively includes unlinked one- and two-body
diagrams (see, e.g., Fig.~10 of Ref.~\cite{ell77}) as well as the
exclusion-enforcing three-body diagrams we want.  This problem, however, is
more pronounced in the $\Q$-box than the $\X$-box since the latter has no
one-body part and far fewer ways to unlink diagrams by exchanging lines (the
horizontal \bb-decay lines are restricted to have incoming neutrons and
outgoing protons). We therefore effectively include only very few unlinked
diagrams by introducing our restrictions in the $\X$-box; the compensating
benefit is a much better account of Pauli exclusion, an important physical
effect. Diagrams such as the one on the left of Fig.~\ref{fig:cancel} result in
large contributions that should not be present in a full calculation.  We
cancel them with the implicit assumption that the canceling contribution from
the figure on the right-hand side is significantly greater than that of typical
third-order diagrams, which we omit.  Eventually, though, this assumption will
have to be tested explicitly.

\section{\label{s:results}Results}

\begin{table}
\centering
\begin{tabular*}{0.48\textwidth}{@{\extracolsep{\fill}}lcc}\hline\hline
& \Ge & \Se \\[.5mm]
\hline
Bare matrix element $M_{0\nu}$& 3.12  & 2.73 \\[.5mm]
First-order $\X$-box, without 3p-1h & 5.44 & 4.86 \\[.5mm]
Full first-order $\X$-box  & 2.20 & 2.40 \\[.5mm]
First order folded & 3.11& 2.79\\[.5mm]
Full second-order $\X$-box & 4.14 & 3.92 \\[.5mm]
\hline
Final matrix element & 3.77 & 3.62\\[.5mm]
\hline
CD-Bonn $G$-matrix & 3.62 & 3.45\\[.5mm]
N$^3$LO $G$-matrix & 3.48 & 3.33 \\[.5mm]\hline\hline
\end{tabular*}
\caption{The \bbz-decay matrix elements $M_{0\nu}$ for \Ge and \Se at various 
approximations in our many-body framework.}
\label{tab:results}
\end{table}

To obtain our final corrected shell-model \bbz-decay matrix elements, we
combine the individual two-body matrix elements of our effective operator with
two-body shell-model transition densities. Since our aim is a consistent
calculation without empirical adjustment, we really ought to take two-body
densities from the diagonalization of a valence-space interaction that is
derived directly from NN+3N forces.  While work in this direction is in
progress, the computation is not yet possible in nuclei this heavy.  Instead we
use two-body densities from existing shell-model calculations, the interactions
for which have been tweaked to fit experimental data in nearby nuclei. For \Ge
we use the calculation of Horoi \cite{hor12} and for \Se that of
Ref.~\cite{men08}; the authors of both have kindly supplied us with their
transition densities.  

Table~\ref{tab:results} presents our matrix elements at various levels of
$\X$-box and folding approximations, using \vlk and taking intermediate-state
excitations to $18\hw$. Despite differences in the NN interaction and size of
the basis space, contributions from first-order diagrams in both \Ge and \Se
largely agree with those first identified in Ref.~\cite{eng09}:
particle-particle and hole-hole ladders together enhance the matrix element,
while the three-particle one-hole diagrams cause a dramatic reduction.  When
folding is included, however, the net correction from first-order $\Q$- and
$\X$-boxes essentially disappears.  Taking the complete set of second-order
diagrams into account, we find a significant enhancement followed by a modest
quenching from folding.  The final matrix element is approximately 20\% percent
larger than the bare matrix element in \Ge and about 30\% larger in \Se.  The
primary reason for the different effects in \Se and \Ge is the difference in
the omitted intermediate-state orbits discussed in Section~\ref{ss:xbox}.  If
we include those orbits, as is standard practice in the construction of
effective interactions, the matrix element is reduced by about 10\% in \Ge and
15\% in \Se.  In Ref.\ \cite{lin01}, which contains a preliminary account of
our calculations in \Se, we obtained 3.56 instead of 3.62.  The small
difference is due to the inclusion in Ref.\ \cite{lin01} of $\Q$-box
restrictions and the addition here of a term in the expansion of the norm
denominator.  Though the two results are close, we believe that the one
reported here is likely closer to the real matrix element.

\begin{table}[t]
\begin{center}
\begin{tabular*}{0.48\textwidth}{ccccccc}
\hline\hline
 \, & \,\, $8\hw$ & \,\, $10\hw$ & \,\, $12\hw$\, & \,\, $14\hw$ & \,\, $16\hw$ & \,\, 
 $18\hw$ \\ [.05in]\hline
Full 1st order & \,\, $2.429$ & \, $2.407$ & \, $2.403$&\, $2.401$ &\, $2.399$&\, 
$2.399$\\[0.5mm]
Full 2nd order & \,\, $3.908$ & \, $3.932$ & \, $3.940$&\, $3.931$ &\, $3.925$&\, 
$3.924$\\[0.5mm] \hline
Final & \,\, $3.489$ & \, $3.553$ & \, $3.595$ & \, $3.611$&\, $3.617$&\, $3.618$
\\[0.5mm]
 
\hline\hline
\end{tabular*}
\vspace*{-2.5mm}
\caption{Convergence of \bbz-decay matrix element in \Se with respect to allowed 
intermediate-state excitations.  In all cases we work in a harmonic-oscillator basis
of 13 major shells.
\label{tab:conv}}
\vspace*{-4mm}
\end{center}
\end{table}               

Several other aspects of the calculation are robust.  As seen in Table
\ref{tab:conv}, our \vlk results at $18\hw$ are well converged to 3 or 4
digits.  And as Table \ref{tab:results} shows, changing the interaction to a
$G$-matrix (in 11 major oscillator shells) in place of \vlk does not affect the
results substantially.  Finally, although we emphasized our procedure of
requiring intermediate-particle lines in $\X$-box diagrams to be unoccupied in
the nucleus in question, other prescriptions yield similar results once norms
and folding are included:  in \Se, for example, we obtain a final matrix
element of 3.50 if we restrict particle lines in both the $\Q$ and $\X$ boxes,
and 3.03 if we impose no restrictions at all.  We should note, however, that at
various intermediate stages of the calculation, the procedures yield quite
different results.  And other parts of the calculation leave room for change as
more physics is included.

\begin{table}[b]
\centering
\begin{tabular}{lcc} \hline\hline
&Restricted & Unrestricted \\[.05in]
\hline
Bare matrix element $M^{\rm cl}_{2\nu}$ & 0.57  & 0.57 \\[0.5mm]
First-order $\X$-box, without 3p-1h & 0.99 & 0.89 \\[0.5mm]
Full first-order $\X$-box  & 0.37 & -0.60 \\[0.5mm]
Full second-order $\X$-box & 0.79 & -0.37 \\[0.5mm]\hline
Final matrix element & 0.96 & 0.70 \\[0.5mm] \hline\hline
\end{tabular}
\caption{The \bbt-decay closure matrix $M_{2\nu}^\text{cl}$ for \Ge at
several levels of approximation, with and without restrictions on occupied
intermediate-particle lines.}
\label{tab:2nu-results}
\end{table}

We turn now to a discussion of \bbt decay.  As noted above, we use the closure
matrix element $M_{2\nu}^\text{cl}$ as a proxy for the full matrix element, a
step that limits how much we can say.  Table \ref{tab:2nu-results} shows the
matrix element for \Ge with and without the intermediate-state restrictions we
impose on occupied or partially occupied orbits in calculating $M_{0\nu}$.
Imposing the restrictions here increases the matrix element, as in \bbz decay;
in this case, however, the result is probably undesirable, given that shell
model calculations of \bbt decay in \Ge overestimate $M_{2\nu}$.  On the other
hand, omitting the restrictions increases the negative contribution of the
3p-1h diagram to such an extent that the matrix element changes sign. The sign
is eventually reversed by higher-order contributions and folding, ultimately
yielding a result that is approximately unchanged from the bare matrix
element.  It is difficult to be comfortable, however, with a low-order
correction that changes the sign of the matrix element.  The sensitivity of the
numbers suggest that terms with more folds, of higher order, or involving more
valence orbitals could also have a significant effect.  

\begin{figure}[t]
\centering
\includegraphics[width=.45\columnwidth]{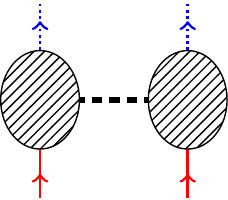}
\caption{\label{fig:blobs} (Color online) Schematic representation of diagrams
contributing to renormalization of $g_A$ in \bbt decay.}
\end{figure}

Another reason (aside from the sign changes in the right-hand column of Table
\ref{tab:2nu-results}) for preferring to restrict intermediate-sate orbits in
the $\X$-box is connected to the long-standing problem of the apparent
suppression of the axial-vector coupling constant $g_A$ in the nuclear medium
\cite{tow87}.  While the suppression probably has many sources, configurations
outside the valence space are likely to play a key role.  Though the bare
operator governing weak decay is one-body, we can simulate the effect of $g_A$
suppression in \bbt decay by including only closure diagrams that have the form
shown in Fig.~\ref{fig:blobs}.  Such diagrams, in which only a single
\bbt-decay line connects the two nucleons, incorporates only the
renormalization of the one-body weak current.  

When we base our calculation of $M_{2\nu}^\text{cl}$ on only these diagrams and
at the same time account for occupied intermediate-state orbits, we find in \Se
that the full result is smaller than the bare result by 38\%, implying an
effective $g_A$ of about 1.0, a reasonable value (in \Ge the value is about
0.7).  On the other hand, when we take no account of occupied orbits we find
(again, with only diagrams of the form in Fig.~\ref{fig:blobs} included) that
the closure matrix element changes sign, something that is impossible through
the renormalization of $g_A$ alone. The sign change reflects the same strong
effect of the 3p-1h diagrams observed in Table \ref{tab:2nu-results}. Though
these considerations are not conclusive, they do indicate that taking the Pauli
principle into account is a beneficial.  Work is currently underway to
investigate $g_A$ quenching more directly.  By focusing on the one-body
operator, we can use analogous many-body techniques, based again on chiral NN
and 3N physics, with the effects of two-body currents implemented consistently
in the bare operator~\cite{men11} to understand the origin of $g_A$ quenching.

Whatever the outcome of that investigation, it is clear that the \bbt matrix
element is sensitive to many details in the wavefunctions, much more sensitive
than its \bbz counterpart.  Thus, although the increase of the \bbt closure
matrix element does not bolster the case for our \bbz calculation, neither, in
our view, does it weaken it much.

\section{\label{s:discussion}Discussion and Outlook}

We have used chiral nuclear forces and many-body perturbation theory to
calculate an effective shell-model \bbz-decay operator, taking into account
corrections to the bare operator from configurations outside the valence to
second order in the interaction.  The resulting nuclear matrix element is
approximately 20\% larger than the bare matrix element in \Ge and about 30\%
larger in \Se.  These new results represent our current best estimates for the
matrix elements but probably do not tell the whole story.  We have omitted a
number of effects that could further alter the results.  To do better, we must
first establish consistency between the Hamiltonian and our effective
operator.  This will require the construction of full non-empirical
valence-space interactions in the $pf$ shell from NN and 3N forces; work in
that direction is in progress.  A related improvement will be to include the
effects of chiral 3N forces in the $\X$-box, in addition to the effects of
chiral two-body currents in the bare operator \cite{men11}.  

At the many-body level, the importance of third- and higher-order terms in the
$\X$-box and additional folding contributions must be understood.  Since we
have found the effects of bubble diagrams to be the most important in our
perturbative expansion, it would be worthwhile to pursue a nonperturbative
calculation of the effects of core polarization (which these diagrams
represent), like that done for effective interactions in Ref.~\cite{hol05}.
Perhaps the most significant obstacle to a truly reliable result, however, is
the implementation of induced three-body operators. Recent work
\cite{shu11,eng04} indicates that such operators are not negligible, and even
here we have shown that three-body diagrams of the form in
Fig.~\ref{fig:cancel} are important.  Unfortunately, the number of induced
three-body diagrams is so large that nobody has computed them even in the
construction of effective interactions.  We must find a way to at least
estimate their size if we want to pursue perturbation theory to its
conclusion.  Controlled nonperturbative approaches \cite{tsukiyama12,jansen11}
are on the horizon, but the inclusion of induced three-body terms is
technically difficult there as well.  In none of these approaches is the
problem impossible to overcome, but doing so will require diligence and
creativity.

\begin{acknowledgments}

We thank M.\ Hjorth-Jensen, M.\ Horoi, J.\ Men\'{e}ndez, and A.\ Poves for
helpful discussions, and Drs.\ Horoi and Poves for providing us with their
shell-model densities.  This work was supported by the BMBF under Contract No.\
06DA70471, the Helmholtz Association through the Helmholtz Alliance Program,
contract HA216/EMMI ``Extremes of Density and Temperature: Cosmic Matter in the
Laboratory'', and the US DOE Grants DE-FC02-07ER41457 (UNEDF SciDAC
Collaboration) and DE-FG02-96ER40963.  J.E.\ gratefully acknowledges in
addition the support of the U.S.\ Department of Energy through Contract No.\
DE-FG02-97ER41019. 

\end{acknowledgments}

\end{document}